# Studies on the switching speed effect of the phase shift keying in SLED for generating high power microwave

Xiong Zhengfeng[1,2](熊正锋)　Cheng Cheng[1](程诚)　Yu Jian[1](于鉴)　Chen Huaibi[1](陈怀璧)　Ning Hui[2](宁辉)

1 Department of Engineering Physics, Tsinghua University, Beijing 100084, China

2 Laboratory on Science and Technology of High Power Microwave, Northwest Institute of Nuclear Technology, Xi'an, Shaanxi 710024, China

**Abstract**: SLAC energy doubler (SLED) type radio-frequency pulse compressors are widely used in large-scale particle accelerators for converting long-duration moderate-power input pulse into short-duration high-power output pulse. The phase shift keying (PSK) is one of the key components in SLED pulse compression systems. Performance of the PSK will influence the output characteristics of SLED, such as rise-time of the output pulse, the maximal peak power gain, and the energy efficiency. In this paper, high power microwave source based on power combining and pulse compression of conventional klystrons was introduced, the nonideal PSK with slow switching speed and without power output during the switching process were investigated, the experimental results with nonideal PSK agreed well with the analytical results.

## 1 Introduction

The high power microwave (HPM) attract increasing attentions due to the applications in radar, communication, heating in fusion, directed energy and microwave undulator[1]. Backward wave oscillator (BWO), magnetically insulated line oscillator (MILO) and virtual cathode oscillator (VCO) are common HPM generators, although the peak power of these devices had achieved multi-gigawatts with tens of nanosecond pulse width, the total energy efficiency of these HPM systems are not high[2].

With the development of particle accelerator technologies, the peak power of conventional klystrons had been reached hundreds megawatts with multi-microseconds pulse width. RF systems based on power combining and pulse compression of moderate-power klystrons are widely used in large-scale accelerators [3]. This technique may be used as a high energy efficiency system for generating HPM and experiments will be carried out to verify the feasibility.

SLED type pulse compressor systems were chosen for simple and compact structure, convenient way for energy extraction through the PSK located at the low power section [4]. SLED had been widely used in accelerators, however we would pay more attention to the rise time, the peak power of output pulse and the energy efficiency than the flat pulse needed in accelerators. Performance of the PSK will influence the characteristics of SLED, the PSK switching phase and time jitter effects had been studied in reference [5] and will not be considered in this paper. The switching speed effect of the PSK on the rise time and peak power gain will be analyzed in our studies.

## 2 Theory

### 2.1 SLED theory with ideal PSK

Conventional SLED type compressors contain a 3dB coupler and two identical high $Q$ storage cavities, as showed in figure 1. When the phase of the input pulse is reserved 180°, the energy stored in the cavities discharged, the emitted field from the cavities and the reflected field added at the output port of 3dB coupler, then forming the compressed high power pulse [6].

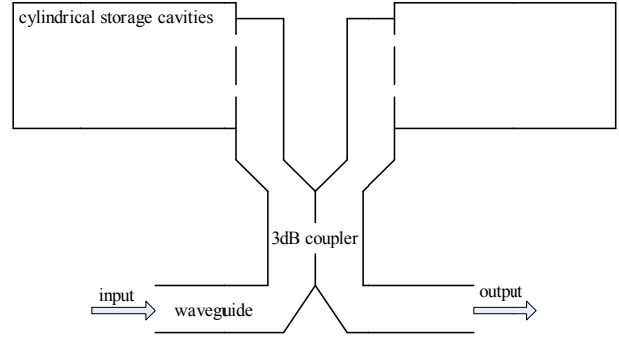

Fig.1. Schematic of the SLED type pulse compressor

By the law of energy conservation and the relations between electric field and power in waveguide, the output field of SLED can be solved by equation [7]:

$$E_{in}^2 = E_{out}^2 + \frac{1}{\beta}E_e^2 + \frac{2Q_0}{\omega\beta}E_e\frac{dE_e}{dt} \quad (1)$$

Where, $\beta$ is the coupling coefficient, $Q_0$ is the unloaded quality factor of the storage cavities. Assuming the reflect coefficient between waveguide and the storage cavities is $\Gamma=-1$, then equation (1) can be expressed as:

$$\tau_c\frac{dE_e}{dt} + E_e = \alpha E_{in} \quad (2)$$

Where, $\tau_c = 2Q_0/\omega(1+\beta)$ is the time constant of storage cavity, $\alpha = 2\beta/(1+\beta)$.

In the ideal case, phase of the input pulse changed instantly when the PSK acted at time $t_1$, as showed in figure 2(a), and the normalized input field in waveguide is showed as figure 2(b).

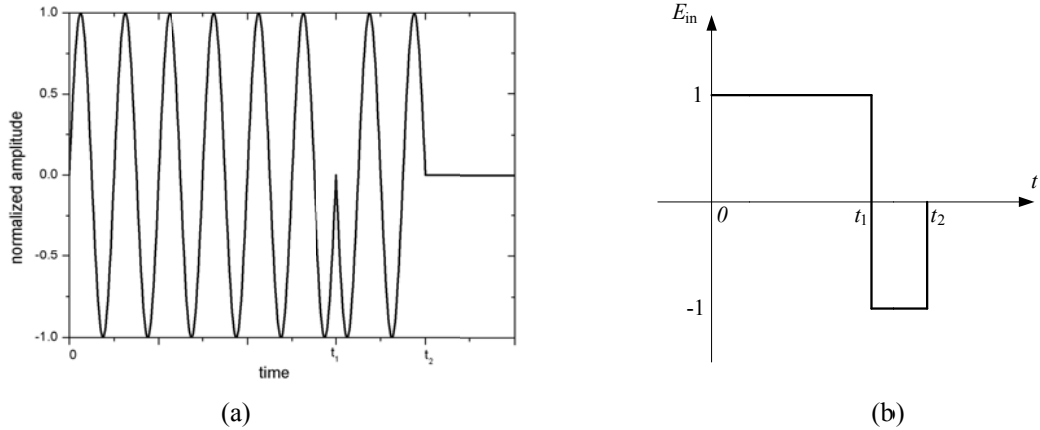

(a)          (b)

Fig.2. Schematic of the input field with ideal PSK

The normalized input field can be expressed as:

$$E_{in}(t) = \begin{cases} 1, & 0 \leq t < t_1 \\ -1, & t_1 \leq t < t_2 \\ 0, & t_2 < t \end{cases} \quad (3)$$

At time instant t=0, there is no emit field from the cavity as $E_e$=0, the solution of equation (2) is:

$$E_e(t) = \begin{cases} \alpha(1-e^{-t/\tau_c}), & 0 \leq t < t_1 \\ \alpha\left[(2-e^{-t_1/\tau_c})e^{-(t-t_1)/\tau_c} - 1\right], & t_1 \leq t < t_2 \\ \alpha\left[(2-e^{-t_1/\tau_c})e^{-(t_2-t_1)/\tau_c} - 1\right]e^{-(t-t_2)/\tau_c}, & t_2 \leq t \end{cases} \quad (4)$$

Then, the normalized output field of SLED:

$$E_{out}(t) = \begin{cases} \alpha(1-e^{-t/\tau_c})-1, & 0 \leq t < t_1 \\ \alpha\left[(2-e^{-t_1/\tau_c})e^{-(t-t_1)/\tau_c} - 1\right]+1, & t_1 \leq t < t_2 \\ \alpha\left[(2-e^{-t_1/\tau_c})e^{-(t_2-t_1)/\tau_c} - 1\right]e^{-(t-t_2)/\tau_c}, & t_2 \leq t \end{cases} \quad (5)$$

Assuming $Q_0=10^5$, $\beta=5$ at frequency 2856 MHz, the input pulse width is 4μs and the phase reversed at the last 1μs, then the normalized field and power of SLED is showed as figure 3. The maximal peak power gain is about 5.5 at the time instant $t_1$ when the phase revered and then decaying exponentially.

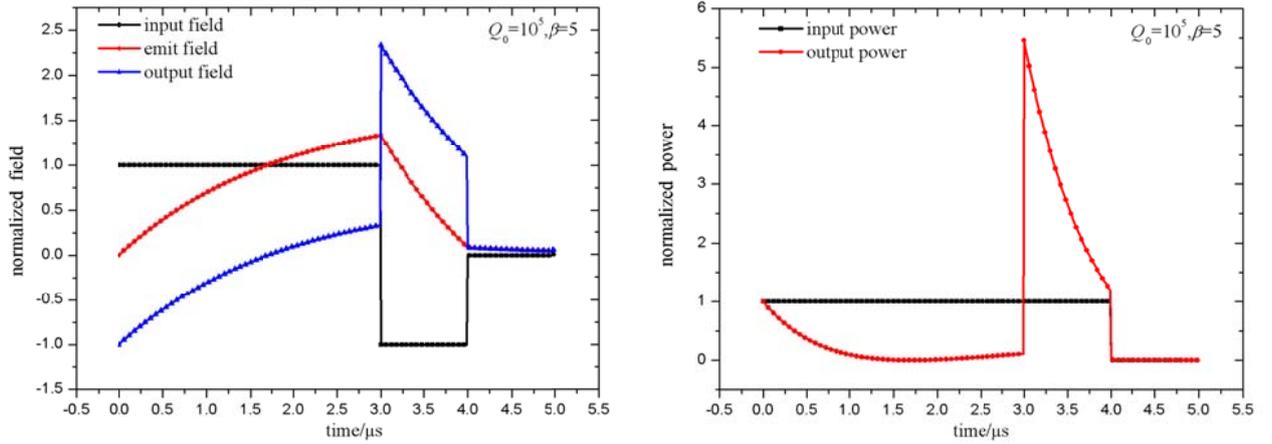

Fig.3. Normalized field and power of SLED with ideal PSK

**2.2 SLED theory with nonideal PSK**

The switching speed of PSK will influence the performance of SLED. Actually, PSK in communication and radar is usually electrically controlled phase shifter. The phase cannot be changed instantly limited by the response time of the device. The actual test results show that the phase shifting process of general PSK will take some time or even without power output during the process.

PSK with slow switching speed will lead to variation of the input field during the phase shifting process, assuming the phase begin to change at time $t_1$ and reversed at time $t_2$, the normalized input field in waveguide is showed in figure 4.

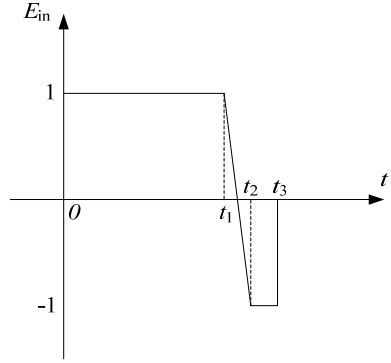

Fig.4. Schematic of the normalized input field with slow PSK

The normalized input field in this case can be expressed as:

$$E_{in}(t) = \begin{cases} 1, & 0 \leq t < t_1 \\ \dfrac{-2t}{t_2 - t_1} + \dfrac{t_2 + t_1}{t_2 - t_1}, & t_1 \leq t < t_2 \\ -1, & t_2 \leq t < t_3 \\ 0, & t_3 \leq t \end{cases} \quad (6)$$

The emitted field from the cavity and the output field can be given as:

$$E_e(t) = \begin{cases} \alpha\left(1 - e^{-t/\tau_c}\right), & 0 \leq t < t_1 \\ \alpha\left[1 - e^{-t_1/\tau_c} + \dfrac{2\tau_c + t_2 + t_1}{t_2 - t_1} - \dfrac{2t}{t_2 - t_1} - \left(1 + \dfrac{2\tau_c}{t_2 - t_1}\right)e^{-(t-t_1)/\tau_c}\right], & t_1 \leq t < t_2 \\ \alpha\left\{\left[1 - e^{-t_1/\tau_c} + \dfrac{2\tau_c}{t_2 - t_1} - \left(1 + \dfrac{2\tau_c}{t_2 - t_1}\right)e^{-(t_2 - t_1)/\tau_c}\right]e^{-(t-t_2)/\tau_c} - 1\right\}, & t_2 \leq t < t_3 \\ \alpha\left\{\left[1 - e^{-t_1/\tau_c} + \dfrac{2\tau_c}{t_2 - t_1} - \left(1 + \dfrac{2\tau_c}{t_2 - t_1}\right)e^{-(t_2 - t_1)/\tau_c}\right]e^{-(t_3 - t_2)/\tau_c} - 1\right\}e^{-(t-t_3)/\tau_c}, & t_3 \leq t \end{cases} \quad (7)$$

$$E_{out}(t) = \begin{cases} \alpha\left(1 - e^{-t/\tau_c}\right) - 1, & 0 \leq t < t_1 \\ \alpha\left[1 - e^{-t_1/\tau_c} + \dfrac{2\tau_c + t_2 + t_1}{t_2 - t_1} - \dfrac{2t}{t_2 - t_1} - \left(1 + \dfrac{2\tau_c}{t_2 - t_1}\right)e^{-(t-t_1)/\tau_c}\right] + \dfrac{2t}{t_2 - t_1} - \dfrac{t_2 + t_1}{t_2 - t_1}, & t_1 \leq t < t_2 \\ \alpha\left\{\left[1 - e^{-t_1/\tau_c} + \dfrac{2\tau_c}{t_2 - t_1} - \left(1 + \dfrac{2\tau_c}{t_2 - t_1}\right)e^{-(t_2 - t_1)/\tau_c}\right]e^{-(t-t_2)/\tau_c} - 1\right\} + 1, & t_2 \leq t < t_3 \\ \alpha\left\{\left[1 - e^{-t_1/\tau_c} + \dfrac{2\tau_c}{t_2 - t_1} - \left(1 + \dfrac{2\tau_c}{t_2 - t_1}\right)e^{-(t_2 - t_1)/\tau_c}\right]e^{-(t_3 - t_2)/\tau_c} - 1\right\}e^{-(t-t_3)/\tau_c}, & t_3 \leq t \end{cases} \quad (8)$$

Figure 5 shows the normalized output field and power with different switching speed of PSK. The duration of phase shifting process is expressed as $\Delta t = t_2 - t_1$. The switching speed has almost no effect on the peak power gain and the maximal peak power also achieved at the phase reversed instant. Longer $\Delta t$ will lead to slower rise time of output pulse and more energy left over after the input pulse finished, it is disadvantageous for generating high power microwave.

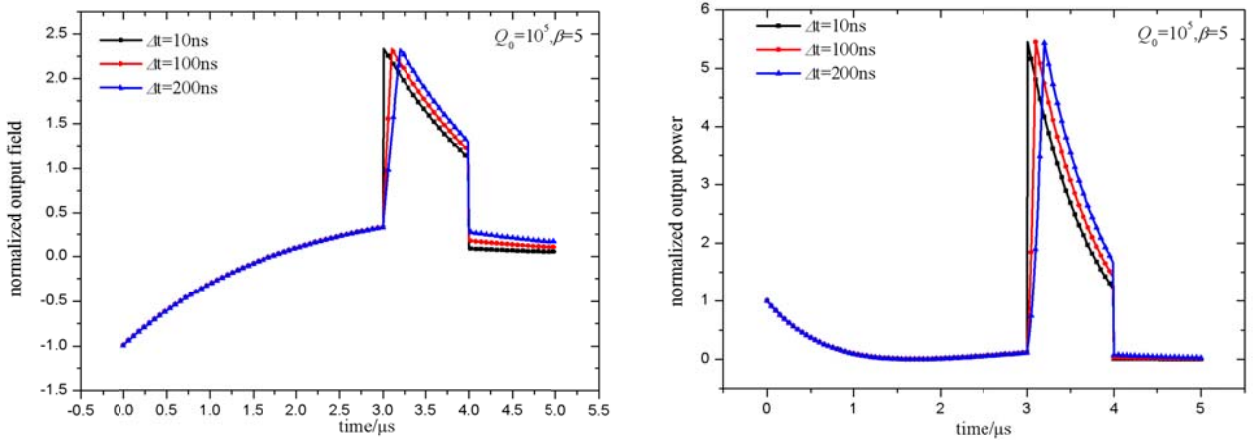

Fig.5. Normalized output field and power of SLED using PSK with different switching speed

Figure 6 shows the normalized input filed in waveguide with the PSK which have no power output during the phase shifting process.

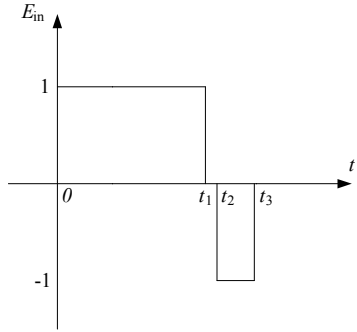

Fig. 6. Schematic of the normalized input field with PSK which have no power output during the phase shifting process

The normalized input field in this case can be expressed as:

$$E_{in}(t) = \begin{cases} 1, & 0 \leq t < t_1 \\ 0, & t_1 \leq t < t_2 \\ -1, & t_2 \leq t < t_3 \\ 0, & t_3 \leq t \end{cases} \quad (9)$$

The emitted field from the cavity and the output field can be given as:

$$E_e(t) = \begin{cases} \alpha\left(1-e^{-t/\tau_c}\right), & 0 \leq t < t_1 \\ \alpha\left(1-e^{-t_1/\tau_c}\right)e^{-(t-t_1)/\tau_c}, & t_1 \leq t < t_2 \\ \alpha\left\{\left[1+\left(1-e^{-t_1/\tau_c}\right)e^{-(t_2-t_1)/\tau_c}\right]e^{-(t-t_2)/\tau_c}-1\right\}, & t_2 \leq t < t_3 \\ \alpha\left\{\left[1+\left(1-e^{-t_1/\tau_c}\right)e^{-(t_2-t_1)/\tau_c}\right]e^{-(t_3-t_2)/\tau_c}-1\right\}e^{-(t-t_3)/\tau_c}, & t_3 \leq t \end{cases} \quad (10)$$

$$E_{out}(t) = \begin{cases} \alpha\left(1-e^{-t/\tau_c}\right)-1, & 0 \leq t < t_1 \\ \alpha\left(1-e^{-t_1/\tau_c}\right)e^{-(t-t_1)/\tau_c}, & t_1 \leq t < t_2 \\ \alpha\left\{\left[1+\left(1-e^{-t_1/\tau_c}\right)e^{-(t_2-t_1)/\tau_c}\right]e^{-(t-t_2)/\tau_c}-1\right\}+1, & t_2 \leq t < t_3 \\ \alpha\left\{\left[\left(1+\left(1-e^{-t_1/\tau_c}\right)e^{-(t_2-t_1)/\tau_c}\right)e^{-(t_3-t_2)/\tau_c}\right]-1\right\}e^{-(t-t_3)/\tau_c}, & t_3 \leq t \end{cases}$$

(11)

Figure 7 shows the normalized field and power of SELD using PSK with $\Delta t$ = 200ns. PSK without power output during the phase shifting process will lead to the front edge of output pulse cutting off and the maximal power gain reducing. The maximal peak power gain is about 4.8 at the time instant $t_2$.

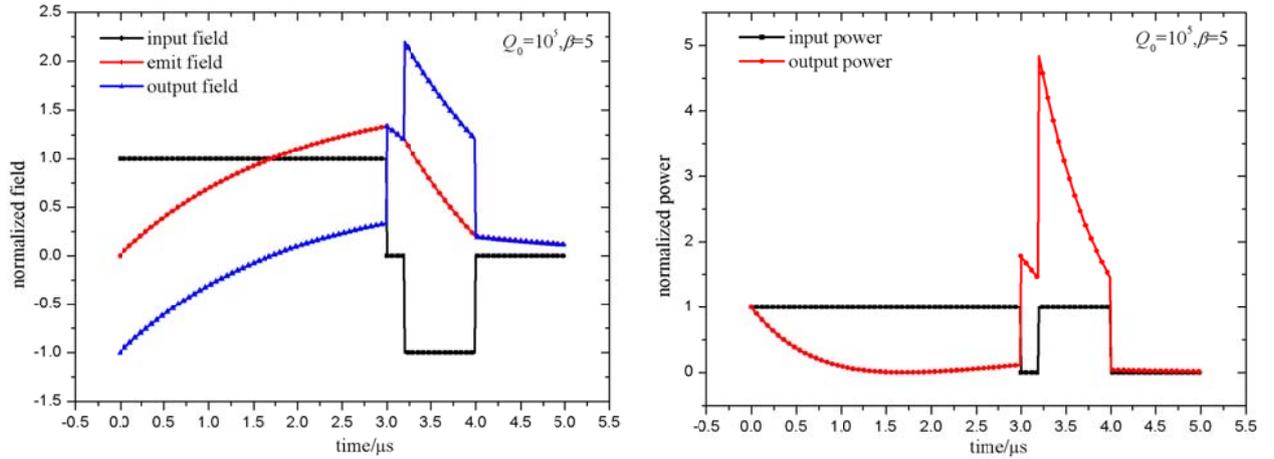

Fig. 7. Normalized field and power of SLED using PSK without power output during the phase shifting process

**3 Experiment**

In the experiments of power combining and pulse compression of conventional klystrons, a PSK module which is integrated in the frequency synthesizer was used. Test result indicated that there is about 200ns without power output during the phase shifting process. Figure 8 shows the envelope waveform of the input pulse (CH3) and the output pulse (CH4), the peak power gain is about 4.7, the front edge of the output pulse was cut off, the waveform of output pulse after the input pulse was finished showed the energy stored in the cavities was not discharged completely.

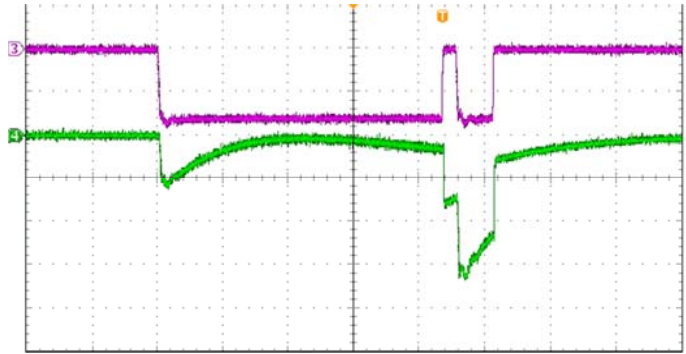

Fig.8. Test waveform of SLED with nonideal PSK

**4. Conclusions**

The switching speed effect of the PSK in SLED for generating HPM was studied in this paper. The PSK with slow switching speed had almost no effect on the peak power gain, but longer phase shifting time would lead to slower rise time of the output pulse. The PSK without power output during the phase shifting process will lead to the front edge of output pulse cutting off and the maximal power gain reducing. Nonideal PSK will cause energy left over after the input pulse finished. In order to generate high peak power and fast rise-time HPM pulse, a PSK with switching speed less than 20ns will be designed and fabricated for our experiments.